\documentclass[prb,twocolumn,superscriptaddress,showpacs,amscd,amsmath,amssymb,verbatim]{revtex4}

\usepackage{graphicx,color}
\usepackage[colorlinks=true,linkcolor=blue,citecolor=blue,urlcolor=blue]{hyperref}
\usepackage{textcomp}
\usepackage[OT1,T1]{fontenc}
\usepackage[squaren, Gray, cdot]{SIunits}
\usepackage{amsmath}
\usepackage{nicefrac}
\usepackage{natbib}
\usepackage{revsymb}
\usepackage{color}
\usepackage{ulem}

\newcommand{\SNIO}{Sr$_3$NiIrO$_6$}
\newcommand{\co}{(Color online) }
\newcommand{\NPD}{neutron powder diffraction}
\newcommand{\etal}{{\it et al.}}

\begin{document}

\title{Magnetic order in the frustrated Ising-like chain compound \SNIO}

\author{E. Lefran\c cois}
\email[Corresponding author: ]{lefrancois@ill.fr}
\affiliation{Institut Laue Langevin, 71 avenue des Martyrs, 38042 Grenoble, France}
\affiliation{Univ. Grenoble Alpes, Institut N\'eel, F-38042 Grenoble, France}
\affiliation{CNRS, Institut N\'eel, F-38042 Grenoble, France}
\author{L. C. Chapon}
\affiliation{Institut Laue Langevin, 71 avenue des Martyrs, 38042 Grenoble, France}
\author{V. Simonet}
\affiliation{Univ. Grenoble Alpes, Institut N\'eel, F-38042 Grenoble, France}
\affiliation{CNRS, Institut N\'eel, F-38042 Grenoble, France}
\author{P. Lejay}
\affiliation{Univ. Grenoble Alpes, Institut N\'eel, F-38042 Grenoble, France}
\affiliation{CNRS, Institut N\'eel, F-38042 Grenoble, France}
\author{D. Khalyavin}
\affiliation{ISIS Facility, STFC, Rutherford Appleton Laboratory, Chilton, Oxfordshire OX11 OQX, United Kingdom}
\author{S. Rayaprol}
\affiliation{UGC-DAE CSR, Mumbai Center, R-5 Shed, BARC, Trombay, Mumbai 400085, India}
\author{E. V. Sampathkumaran}
\affiliation{Tata Institute of Fundamental Research, Homi Bhabha Road, Colaba, Mumbai 400005, India}
\author{R. Ballou}
\affiliation{Univ. Grenoble Alpes, Institut N\'eel, F-38042 Grenoble, France}
\affiliation{CNRS, Institut N\'eel, F-38042 Grenoble, France}
\author{D. T. Adroja}
\email[Corresponding author: ]{devashibhai.adroja@stfc.ac.uk}
\affiliation{ISIS Facility, STFC, Rutherford Appleton Laboratory, Chilton, Oxfordshire OX11 OQX, United Kingdom}
\affiliation{Physics Department, University of Johannesburg, PO Box 524, Auckland Park 2006, South Africa}

\date{\today}

\begin{abstract}

We have studied the field and temperature dependence of the magnetization of single crystals of \SNIO. These measurements evidence the presence of an easy axis of anisotropy and two anomalies in the magnetic susceptibility. Neutron powder diffraction realized on a polycrystalline sample reveals the emergence of magnetic reflections below 75 $\kelvin$ with magnetic propagation vector {\bf k} $\sim$ (0, 0, 1), undetected in previous neutron studies [T.N. Nguyen and H.-C zur Loye, J. Solid State Chem., {\bf 117}, 300 (1995)]. The nature of the magnetic ground state, and the presence of two anomalies common to this family of material, are discussed on the basis of the results obtained by neutron diffraction, magnetization measurements, and symmetry arguments. 

\end{abstract}

\pacs{75.25.-j, 75.30.Cr, 75.30.Gw, 75.47.Lx}
\maketitle


\section{Introduction}

\par In the last decades, the oxides of the family A$_3$MM'O$_6$ (A = alkaline-earth metal, M,M' = transition metal) attracted a lot of attention because of their unconventional magnetic properties due to the interplay between low dimensionality, magnetic frustration and magnetocrystalline anisotropy. In these compounds, the magnetic sublattice is usually described as chains arranged on a triangular lattice. The M and M' ions, on trigonal and octahedral sites respectively, alternate on each chain and can interact via the intrachain interaction and several interchain couplings \cite{Chapon2009}. In the Heisenberg classical limit, the magnetic ground state for nearest-neighbor antiferromagnetic interchain interaction on a triangular lattice is the 120$\degree\ $ spin configuration. More generally in the present systems, the ground state is an incommensurate structure with a pitch determined by the relative strengths of the intrachain and interchain interactions \cite{Chapon2009}. In the Ising limit however, the frustration is not released and magnetic structures with modulated moments may be stabilized at finite temperature.  In particular, in the J$_1$/J$_2$ model on a triangular lattice, a partially disordered magnetic configuration system with only two thirds of the chains that order simultaneously has been predicted \cite{Mekata1977} and observed experimentally in CsCoCl$_3$ for instance \cite{Mekata1978}. The most studied compound in the frustrated regime is Ca$_3$Co$_2$O$_6$ because of its large easy-axis anisotropy which confines the moment along the chain axis.  This system famously displays steps in the magnetization below a certain freezing temperature of 10 $\kelvin$. Although the origin of these steps is still a matter of debate, it is certainly a direct consequence of the strong anisotropy  which stabilizes two competing magnetic states, an amplitude modulated phase with wave-vector ${\bf k} = (0,~0,~1.01)$ and a commensurate antiferromagnetic phase \cite{Agrestini2011}.
\par More recently, the attention has been directed towards analog compounds containing transition metal ions of the 4d and 5d series, with the objective of studying the unconventional magnetic properties of these frustrated systems in the strong spin-orbit regime. Magnetic measurements show that Ca$_3$CoRhO$_6$ \cite{Niitaka1999}, Sr$_3$CoIrO$_6$ \cite{Mikhailova2012}, Sr$_3$NiRhO$_6$ \cite{Mohapatra2007}  and \SNIO\ \cite{Nguyen1995,Flahaut2003,Mikhailova2012} have a similar behavior to Ca$_3$Co$_2$O$_6$ with two magnetic anomalies at $T_1$ (90, 90, 45 and 80~$\kelvin$ respectively) and $T_2$ (35, 25, 15 and 15~$\kelvin$ respectively), revealed in the magnetic susceptibility. Below $T_1$, a deviation from the Curie-Weiss behavior is observed followed by a steep increase in the susceptibility and the onset of strong spin dynamics. This dynamic gets frozen at $T_2$. The susceptibility measured after a zero-field cooling is much smaller below this temperature than after a finite field cooling. Whilst the transition temperatures are considerably higher than for Ca$_3$Co$_2$O$_6$, it appears that the magnetic behavior does not depend on the presence of one or two magnetic ions. Indeed in Ca$_3$Co$_2$O$_6$, only one of the Co ions along the chain is magnetic \cite{Sampath2004,Tabuko2005} contrary to the other compounds in which the presence of two alternating magnetic ions leads to more complex exchange interactions along and between the chains. Besides, one would expect a gradual change from localized magnetism in the case of insulating 3d compounds to a magnetic behavior reminiscent of metals for the case of compounds containing 5d transition metal ions. It is therefore relatively surprising to find such similarities in the series. It has been recently shown from ab-initio calculations that for the Ir-based compound \SNIO, the strong spin-orbit coupling opens a gap in the electronic spectrum \cite{Zhang2010,Sarkar2010,Ou2014} that may explain the localized magnetism even in the 5d case.  This result seems corroborated by neutron powder diffraction experiments on Ca$_3$CoRhO$_6$ and Sr$_3$CoIrO$_6$, which evidenced long-range magnetic ordering  below T$_1$ \cite{Niitaka2001,Mikhailova2012} with localized moments on the Co and Rh (Ir) sites. In both cases, the magnetic structure with wave-vector  {\bf k}~=~(0,~0,~1) is compatible either with a partially disordered antiferromagnetic state, i.e. two-thirds of the chains are ordered and antiferromagnetically coupled while one-third remains disordered, or with an amplitude-modulated antiferromagnetic structure.\\
\indent In this article, we study the magnetic properties of the \SNIO\ compound. We report magnetization measurements on single crystals to probe in details the magnetocrystalline anisotropy by applying magnetic field along different crystallographic directions, which was impossible to quantify from previous data obtained on polycrystalline materials. The nature of the magnetic ground state is investigated from neutron powder diffraction measurements as a function of temperature. We propose a model for the magnetic structure compatible with symmetry analysis and discuss the results obtained in comparison with the other compounds of the same family (M = Ni, Co and M' = Rh, Ir, Co).


\section{Experimental}

\par Single crystals of \SNIO\ were grown using the flux method previously described for the isostructural compound Sr$_3$NiPtO$_6$ \cite{Nguyen1994}. The high purity starting materials were SrCO$_3$, NiO, Ir and K$_2$CO$_3$ as solvent. The mixture was heated under air at a rate of 60~$\celsius\per\hour$ to 1050~$\celsius$, held at that temperature for 72~h and ramped to 880~$\celsius$ at 6~$\celsius\per\hour$. The furnace was then naturally cooled down to room temperature. Crystals of hexagonal shape with typical dimension of 0.8x0.8x0.3~$\milli\cubic\meter$, were separated from the remaining K$_2$CO$_3$ flux using distilled water. Powder samples were also prepared by solid state reaction in order to perform \NPD. Details of polycrystalline sample preparation were similar to that given in Ref.~\onlinecite{Mikhailova2012}\\


\par The temperature and field dependence of the single-crystal magnetization were measured using a Quantum Design MPMS\textsuperscript{\textregistered} SQUID magnetometer. The temperature dependence was measured both in zero-field-cooled (ZFC) and field-cooled (FC) mode between 2 and 300 $\kelvin$. In the ZFC protocol, the sample is cooled down to 2~$\kelvin$ in zero magnetic field and the magnetization is measured in $0.01\ \tesla$ while heating. In the FC protocol, the sample is cooled down in $0.01\ \tesla$. The magnetization field dependence was measured in a magnetic field up to 5~$\tesla$.\\


\par Neutron powder diffraction experiments were performed on polycrystalline sample of \SNIO\ at the ISIS pulsed neutron source of the Rutherford Appleton Laboratory, U.K. The powder was loaded into a 6 mm diameter vanadium can inserted in a cryostat. High-resolution data were collected on heating between 2 and 100 $\kelvin$, on the WISH diffractometer on the Target Station 2. Rietveld refinements were carried out using the FULLPROF program \cite{Rodriguez-Carvajal1993}.


\section{Results}

\subsection{Crystal structure of \SNIO}

\begin{figure}[h]
	\centering
	\includegraphics[scale =1, trim = 0.3cm 9cm 12.3cm 15cm, clip]{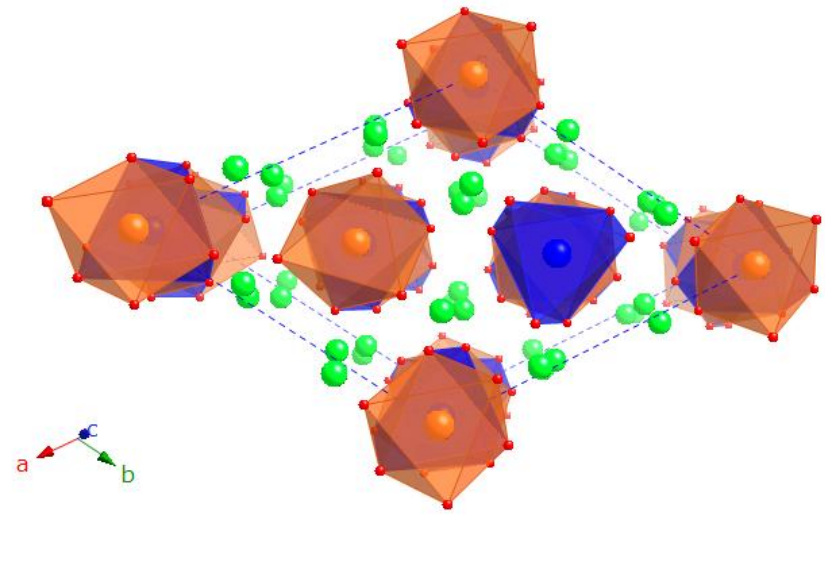} 
	\includegraphics[scale =1, trim = 0.3cm 9.8cm 12cm 13cm, clip]{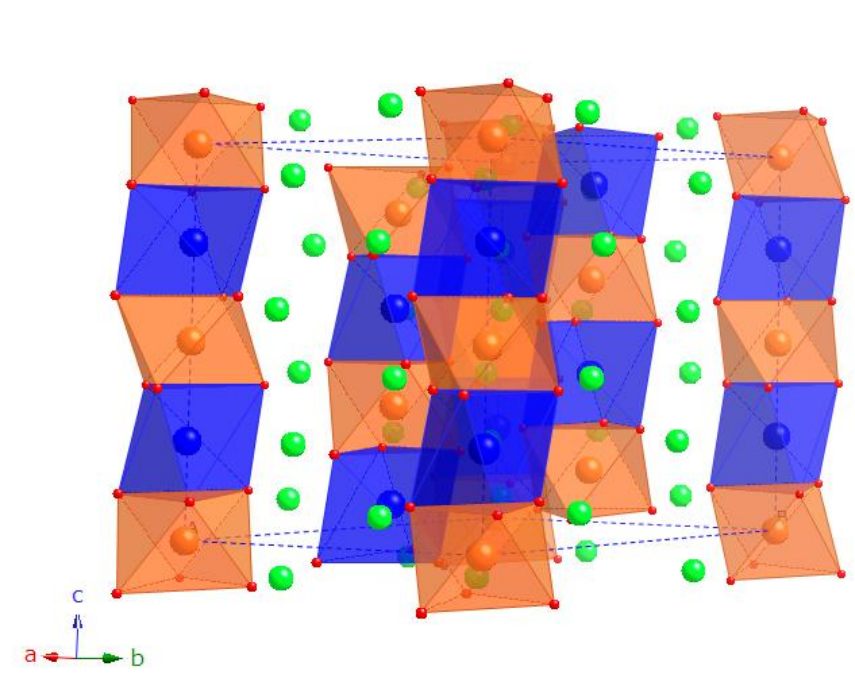} 
	\caption{\co Structure of \SNIO\ shown projected on the (${\bf ab}$) plane (top) and in a perspective view (bottom). Ir$^{4+}$ atoms are represented in orange, Ni$^{2+}$ in blue, Sr$^{2+}$ in green and O$^{2-}$ in red. The IrO$_6$ octahedra and NiO$_6$ trigonal prisms are represented in orange and blue respectively. The conventional hexagonal unit-cell is shown with dotted blue lines.}
	\label{fig:structure}
\end{figure}

\SNIO\ crystallizes in the space group $R\bar{3}c$. The lattice parameters refined from our neutron data at 100 $\kelvin$, $a = 9.6095(2)\ \angstrom$ and $c = 11.1654(8)\ \angstrom$, are consistent with previous results \cite{Nguyen1995}. The crystal structure (see Fig.~\ref{fig:structure}) consists of chains aligned along the ${\bf c}$ direction, formed by alternating face-sharing NiO$_6$ trigonal prisms and IrO$_6$ octahedra. The chains are distributed on a triangular lattice in the (${\bf ab}$) plane. The Ir atom occupies the Wyckoff site 6b (0, 0, 0), i.e. there are two Ir in the primitive unit at (0, 0, 0) and (0, 0, $\frac{1}{2}$). The Ni atom occupies the Wyckoff site 6a (0, 0, $\frac{1}{4}$), generating two  Ni in the primitive unit at (0, 0, $\frac{1}{4}$) and (0, 0, $\frac{3}{4}$). Taking into account the R-lattice translations ${\bf t_1} = (\frac{2}{3}, \frac{1}{3}, \frac{1}{3})$ , ${\bf t_2} = (\frac{1}{3}, \frac{2}{3}, \frac{2}{3})$, and identity, there are six Ir and six Ni ions which occupy each distinct atomic positions in the conventional hexagonal unit-cell. The aforementioned symmetry leads to a shift of $\frac{1}{6}c$ between neighboring chains. 

\par The Ni$^{2+}$ (S = 1) and Ir$^{4+}$ (J$_{eff} = 1/2$) ions are magnetically coupled via several intra- and interchains magnetic interactions. The nearest neighbor intrachain interaction is a superexchange interaction via one oxygen, while the interchain interactions are super-superexchange interactions mediated by two oxygens.

\subsection{Magnetization measurements}

\begin{figure}[h]
	\centering
	\includegraphics[scale=.33, trim = 0.0cm 0.5cm 0.5cm 0.5cm, clip]{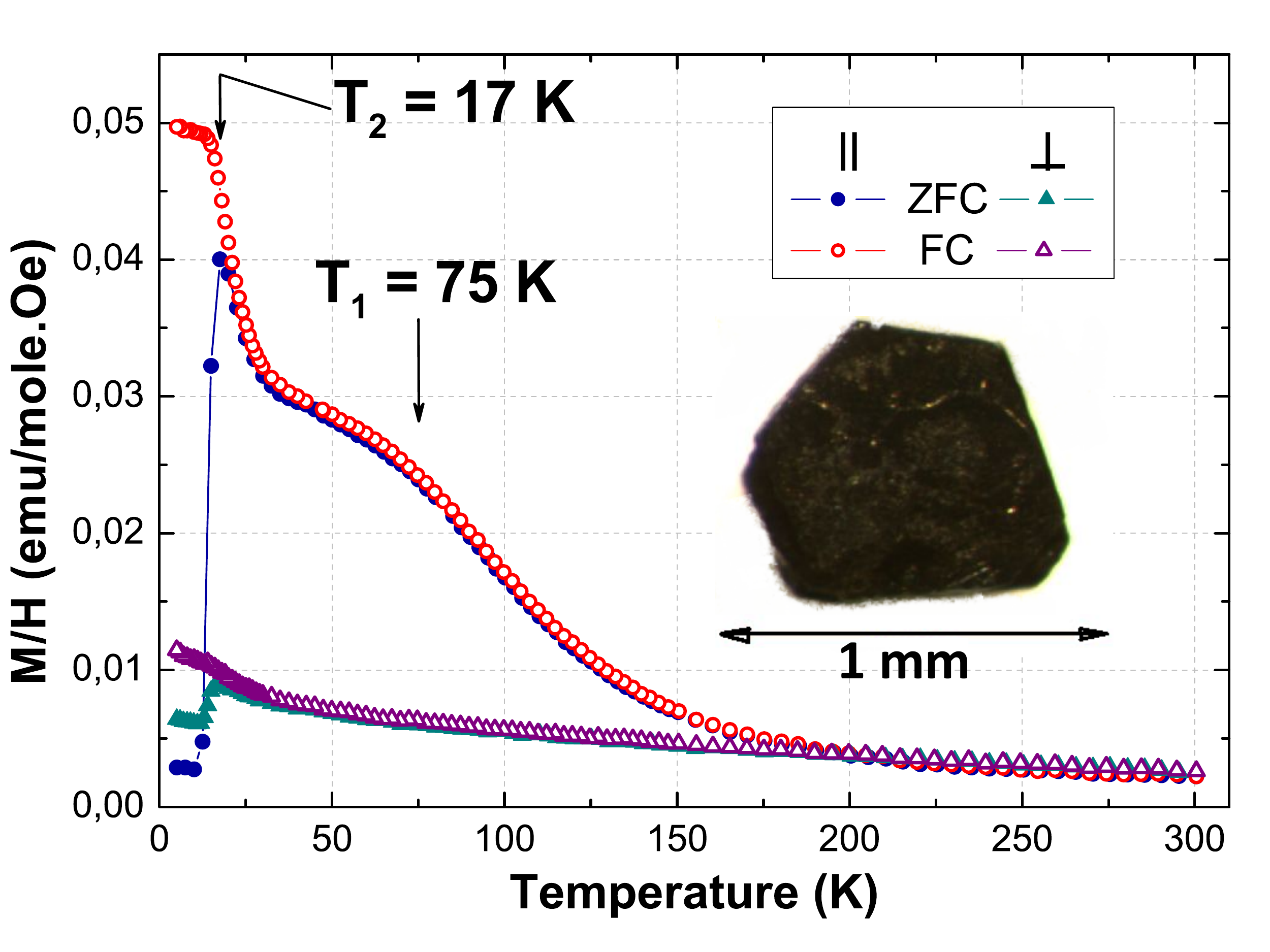}
	\caption{\co Temperature dependence of the magnetic susceptibility measured after a ZFC (filled symbol) and a FC in 0.01 $\tesla$ (empty symbol) with the magnetic field applied parallel and perpendicular to the {\bf c} axis. Inset : picture of a \SNIO\ crystal.}
	\label{fig:chi_t}
\end{figure}

The magnetization measurements were performed on the single crystal with the magnetic field {\bf H} applied parallel ($\parallel$) and perpendicular ($\perp$) to the ${\bf c}$ axis. The temperature dependence of the magnetic susceptibility measured in both directions after a ZFC and a FC in 0.01~$\tesla$ is shown in Fig.~\ref{fig:chi_t}. A significant difference is observed below 200~$\kelvin$: the magnetization is much higher when measured with {\bf H} parallel to {\bf c}  than with {\bf H} perpendicular to {\bf c}. This evidences an easy-axis anisotropy along {\bf c}.  For the measurement with {\bf H} $\parallel$ {\bf c}, a broad maximum is observed around $T_1 = 75\ \kelvin$. Below $T_2 = 17\ \kelvin$, the ZFC and FC curves get separated with a sudden decrease in the magnetic susceptibility after a ZFC.\\

\begin{figure}[h]
	\centering
	\includegraphics[scale=.3]{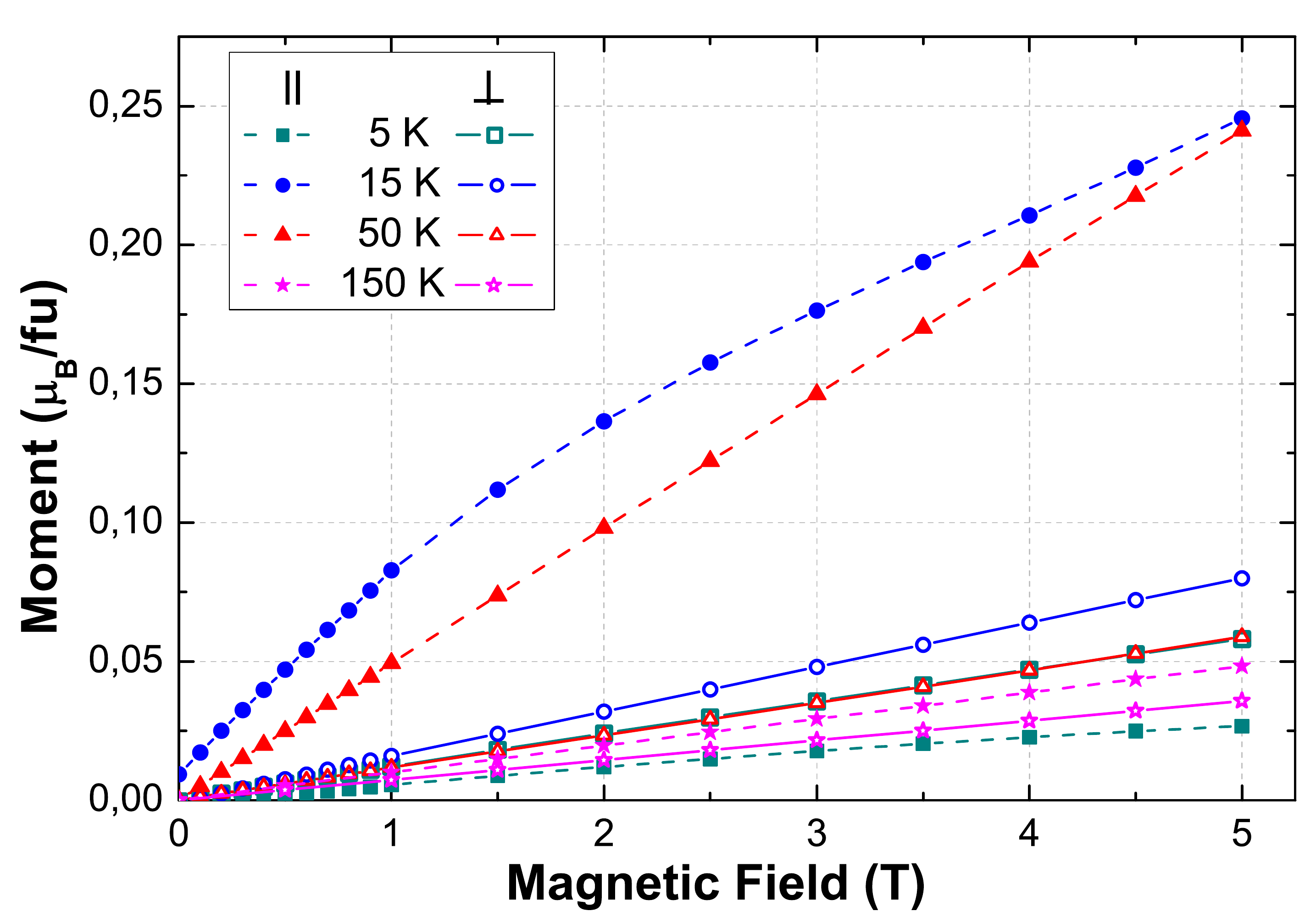}
	\caption{\co Field dependence of the magnetization of \SNIO\ measured at 5 (in cyan), 15 (in dark blue), 50 (in red) and 150 (in purple) $\kelvin$.}
	\label{fig:m_h}
\end{figure}

\begin{figure}[h]
	\centering
	\includegraphics[scale=.3]{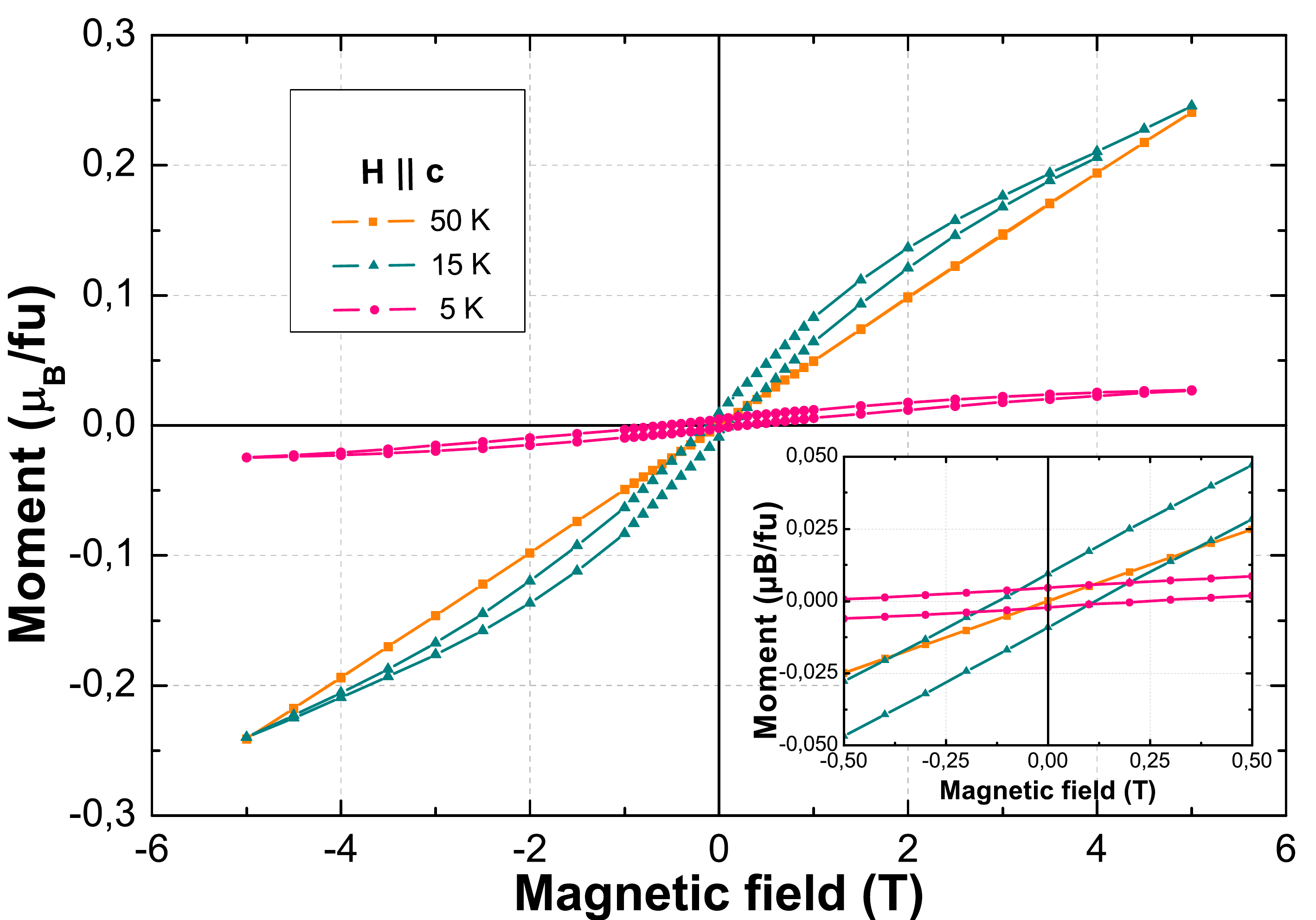}
	\caption{\co Magnetization hysteresis curves measured at 5, 15 and 50~$\kelvin$\ with the magnetic field applied parallel to {\bf c}. Inset : zoom of the hysteresis curves showing the evolution of the hysteresis opening with the temperature in zero field.}
	\label{fig:cycle}
\end{figure}

The field dependence of the magnetization was measured at 5, 15, 50 and 150~$\kelvin$ in a magnetic field up to 5~$\tesla$ applied both perpendicular and parallel to the {\bf c} axis. The results, shown Fig.~\ref{fig:m_h}, are consistent with an easy axis of anisotropy along {\bf c}. A non linearity of the magnetization isotherm manifests itself at 15~$\kelvin$ for {\bf H} $\parallel$ {\bf c}. At 5~$\tesla$ under these conditions, the magnetization reaches a value of $0.25\ \mu_B/f.u$. This announces a step in the magnetization process which was observed by Flahaut \etal\ in high field magnetization measurements \cite{Flahaut2003}. In Co compounds with ferromagnetic intrachain interactions, this magnetization step corresponds to 1/3 of the saturated magnetization \cite{Maignan2000,Niitaka2001a}. In the Ni compounds, the step value can be attributed to 1/3 of the magnetization of ferrimagnetic chains (antiferromagnetic intrachain interactions) which is lower than the total saturated magnetization. For lower temperature, the magnetization decreases significantly while becoming more linear. Figure \ref{fig:cycle} shows the hysteresis cycles measured at 5, 15 and 50 $\kelvin$. While no hysteresis is observed between $T_1$ and $T_2$, the hysteresis cycle opens below T$_2$. It is impossible to extract a value of the ferromagnetic moment below this temperature because one cannot saturate the magnetization cycle
. Its origin will be discussed in a following section of the paper in the lights of neutron magnetic diffraction and symmetry arguments.

\subsection{Neutron powder diffraction}

\begin{figure}[h]
	\centering
	\includegraphics[scale=.3]{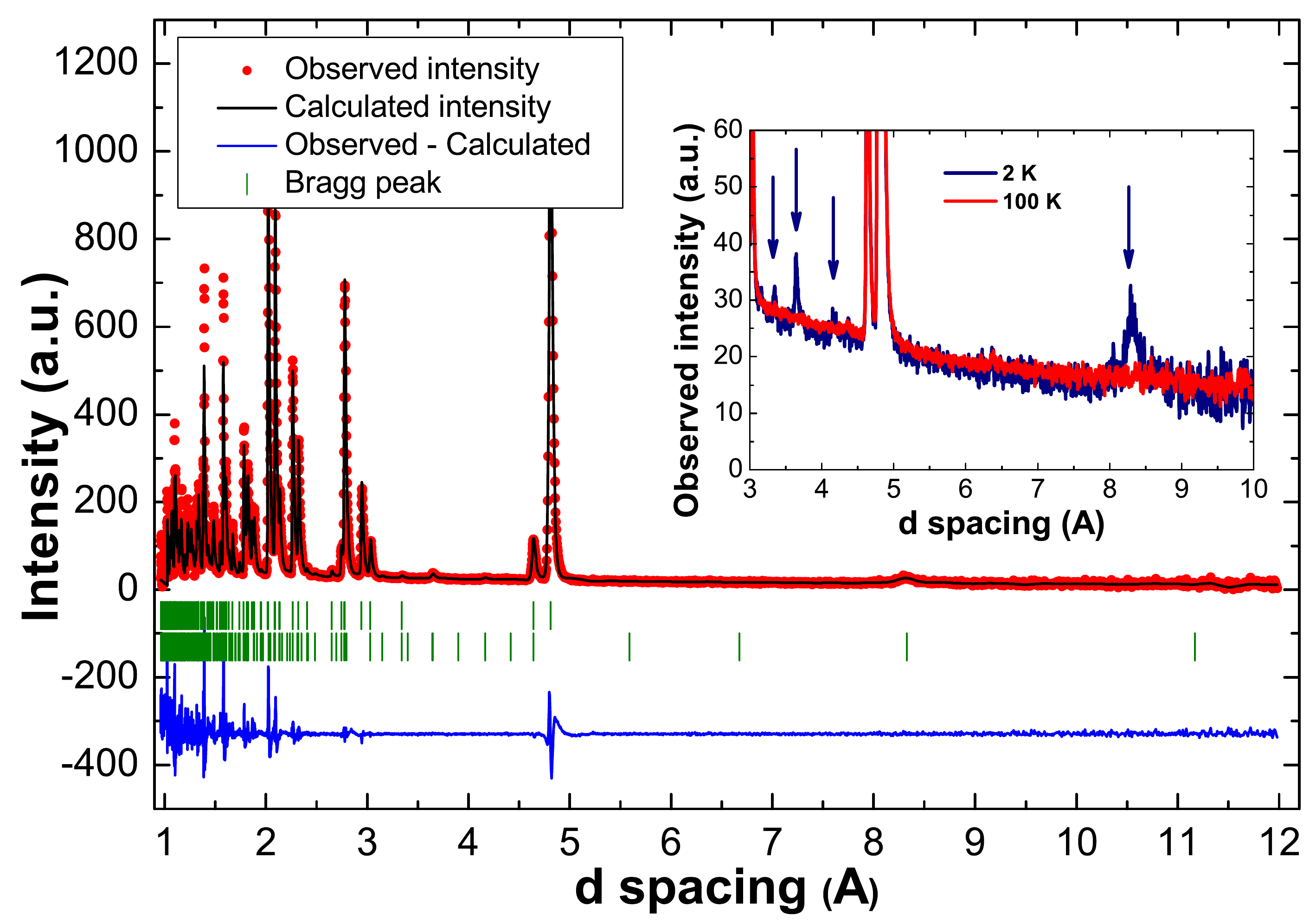}
	\caption{\co Neutron powder diffraction of \SNIO\ recorded at 2 $\kelvin$. The observed and calculated intensities are shown as well as the difference. Inset: zoom of the neutron powder diffraction at 2 and 100 $\kelvin$ showing the onset of magnetic Bragg peaks.}
	\label{fig:2K}
\end{figure}

\begin{figure}[h]
	\centering
	\includegraphics[scale=.3, trim = 0cm 0cm 0.5cm 0cm, clip]{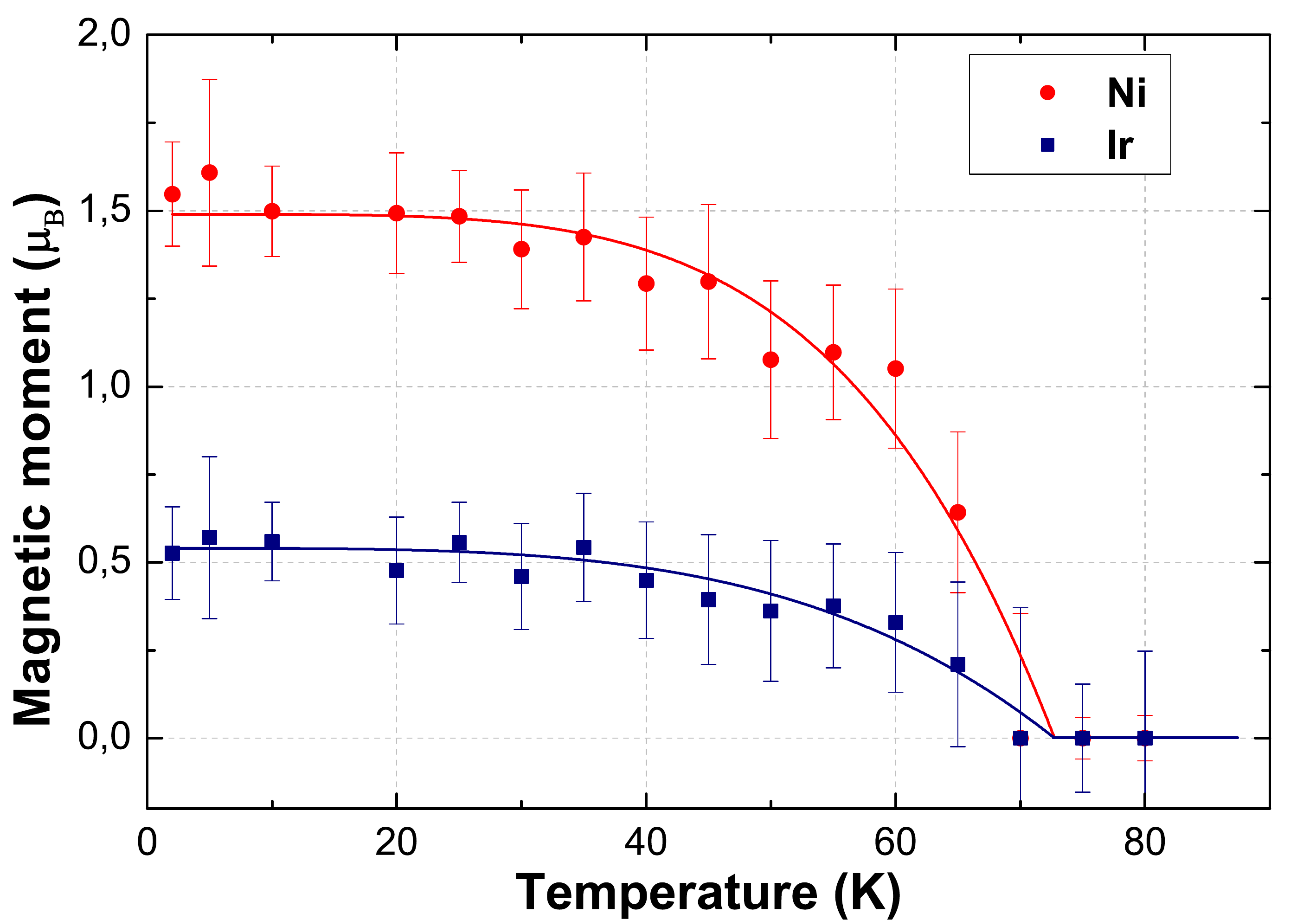}
	\caption{\co Temperature dependence of the magnetic moments of Ni and Ir in absolute value from the refinement of the diffractograms. The plain line are guides for the eyes.}
	\label{fig:mag}
\end{figure}

\begin{figure}[h]
	\centering
	\includegraphics[scale =.35, clip]{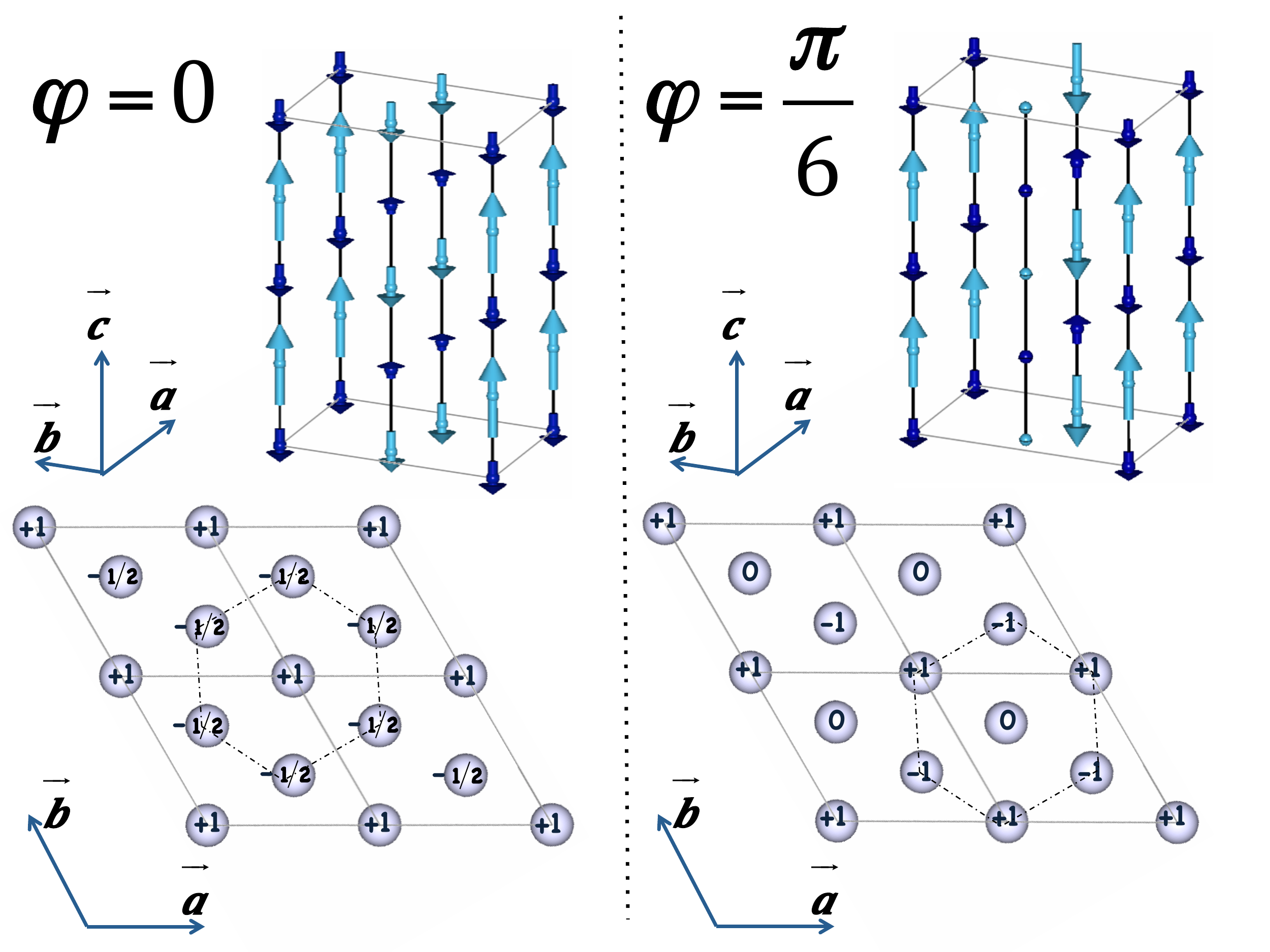} 
\caption{\co Possible magnetic structures of \SNIO\ stabilized at low temperature in a commensurate phase with propagation vector ${\bf k} = (0, 0, 1)$. The structures correspond to two different symmetries dictated by the value of the overall phase $\varphi$ of the modulation, and displayed as perspective views (top) and as sketches projected into the \textit{ab}-plane (bottom). The dark blue (resp. light blue) arrows indicate the magnetic moments on the Ir (resp. Ni) sites. The dashed hexagon in each structure is a guide to the eyes to highlight the transformation properties of the magnetic state under inversion symmetry (see text for details). (Left) Structure of symmetry P$\bar{3}$c'1  corresponding to $\varphi = 0$. (Right) Structure of symmetry P$\bar{3}'$c'1  corresponding to $\varphi = \frac{\pi}{6}$. }
	\label{fig:mag-s}
\end{figure}

A model for the magnetic structure of \SNIO\ has been derived from Rietveld refinement of the neutron powder diffraction collected at 2 $\kelvin$. Below 75 $\kelvin$, 4 additional peaks in the diffraction pattern are observed which are resolution limited (see Fig.~\ref{fig:2K}) and that can be indexed with the commensurate $(0, 0, 0.997(4))$ propagation vector. The intensities of the magnetic peaks saturate below 35 $\kelvin$, i.e. there is no sign of change in the magnetic structure through the temperature $T_2$ at which the ZFC susceptibility suddenly drops (see Fig.~\ref{fig:chi_t} and \ref{fig:mag}). The magnetic form factor of the Ir$^{4+}$ used for the refinement is the one recently determined by Kobayashi {\it et al.} \cite{Kobayashi2011}. The possible magnetic structures compatible with the $R\bar{3}c$ symmetry of \SNIO\ and a second order phase transition were determined using the method of representation analysis \cite{Bertaut1968}. In the group of the wave-vector (little group), there are three irreducible representations, two of dimension one ($\tau_1$, $\tau_2$), corresponding respectively to antiferromagnetic and ferromagnetic arrangements within each magnetic sublattice (Ni and Ir). The third one, of dimension two, represents magnetic ordering in the plane. These modes are listed in Table~\ref{tab:mag-struct}. 
Only the magnetic structure spanned by the irreducible representation $\tau_2$ (in the Kovalev setting \cite{1993representation}, see Table \ref{tab:mag-struct}) is compatible with the experimental data (see Fig.~\ref{fig:mag-s}). It consists of antiferromagnetically coupled Ni$^{2+}$ ($ M_{Ni} = 1.5(1)\ \mu_B$) and Ir$^{4+}$ ($ M_{Ir} = 0.5(1)\ \mu_B$) magnetic moments along the chains. M$_{Ni}$ and M$_{Ir}$ represents the amplitude of the magnetic modulation at 2~$\kelvin$. However, as for any diffraction experiment, the intensities are sensitive only to the modulus of the magnetic structure factor (the Fourier transform of the magnetization), and we do not have access to the global phase of the magnetic modulation, $\varphi$. Therefore the magnetic structure derived from the neutron diffraction data can either be described as an amplitude modulation of the moment for an arbitrary value of $\varphi$, or as  a partially disordered antiferromagnetic state where 2/3 of the chains are antiferromagnetically coupled and the remaining 1/3 stay disordered for $\varphi=\frac{\pi}{6}+n\cdot\frac{2\pi}{3}$ ($n$: integer) as shown Fig.~\ref{fig:mag-s}. Any other value of the phase leads to magnetic structures with non-constant moment. In particular for $\varphi=0$ we get (+M, $-\frac{1}{2}M$, $-\frac{1}{2}M)$ for the moments at atomic positions $\left[(0,0,z),(\frac{1}{3},\frac{2}{3},\frac{2}{3}+z),(\frac{2}{3},\frac{1}{3},\frac{1}{3}+z)\right]$ (see Fig.~\ref{fig:mag-s}). 

More insight into the exact nature of the magnetic ground state can be obtained by combining the neutron results with magnetization measurements. In the previous section, we showed that a small magnetic hysteresis opens below T$_2$ along the c-direction, indicating the presence of a weak ferromagnetic component. The onset of such term appears at first counter-intuitive since the wave-vector {\bf k}=(0, 0, $k$) is inside the Brillouin zone, i.e. the primary order parameter forbids a macroscopic magnetization. However, such mode can exist as a secondary order parameter, its coupling with the primary one imposing strong restrictions on the nature of the magnetic state. The full magnetic symmetry is captured by considering the complete representation LD1 (notation of Miller-Love \cite{Cracknell1979} ) of the paramagnetic group $R\bar{3}c1'$ with {\bf k}=(0, 0, $k$), which corresponds to $\tau_2$ in the little group. Since $\tau_2$ is one-dimensional, LD1 is two dimensional with the matrices \cite{Aroyo2006} given in Tab. \ref{tab:IR} for the group generators acting on the space formed by ($\rho$,$\rho^*$) where $\rho$ is the primary (complex) order parameter. The table also shows how a ferromagnetic component (M$_z$) transforms under the different symmetry operations. An incommensurate structure belongs to the magnetic space group $R\bar{3}c1'$ since time reversal is equivalent to a lattice translation. In such a case, a ferromagnetic component is forbidden. In contrast, when the structure is commensurate with $k$=1, different symmetries are realized depending on the direction of the order-parameter. When $\rho$ is real, the inversion center exists combined with time-reversal symmetry and the magnetic space group is $P\bar{3}'c'1$ prohibiting as well a ferromagnetic component. The corresponding magnetic structure is the one detailed previously with $\varphi=\frac{\pi}{6}$ and shown in Fig. \ref{fig:mag-s}. For $\rho$ purely imaginary, the magnetic space group is $P\bar{3}c'1$ (inversion symmetry is preserved) allowing thus the presence of a ferromagnetic component. This corresponds to the magnetic structure with $\varphi$=0 shown in Fig. \ref{fig:mag-s}. For an arbitrary direction of $\rho$, the inversion symmetry is lost, and the space group is $P3c'1$, also allowing ferromagnetism but with a complex magnetic state. This symmetry analysis shows that the most symmetric magnetic state below T$_2$ compatible with a ferromagnetic component is a commensurate amplitude modulated structure with $\varphi=0$. The onset of a weak ferromagnetic component is explained by a coupling term proportional to $|\rho|^3 \cdot M_z$ in the free energy expansion. The transition above T$_2$ is either a commensurate to incommensurate transition or, if the modulation remains commensurate, an abrupt change of the phase from $\varphi=0$ to $\varphi=\frac{\pi}{6}$. The momentum resolution of our experiment does not allow to probe a possible small incommensurability. It is worth noting that the aforementioned symmetry analysis is valid for all materials of this class ordering magnetically with a wave-vector along the LD (0, 0, $k$) line of symmetry. \\ 
The obtained magnetic structure is clearly compatible with the hypothesis of an easy axis of anisotropy along the {\bf c} axis suggested by the magnetization measurements. The amplitude of the Ni magnetic modulation ($ M_{Ni} = 1.5(1)\ \mu_B$) obtained from the refinement of the diffraction data (Fig.~\ref{fig:mag}) is slightly reduced with respect to the pure spin moment but close to the full magnetic moment calculated by DFT (spin component of 1.69 and orbital component of 0.2 $\mu_B$) \cite{Ou2014}. The amplitude modulation of 0.5(1) $\mu_B$  for the Ir$^{4+}$ sublattice is smaller than the full 1 $\mu_B$ expected for a pure J$_{eff}$ = 1/2 state, indicating a strong contribution from the trigonal crystal field ($\Delta$) which elongates the IrO6 octahedra along the c-axis. It is slightly smaller than that predicted by DFT calculations (0.86 $\mu_B$) \cite{Ou2014}, and compatible with a non negligible value of $\Delta$ (0.21 eV)  with respect to the spin-orbit coupling term (0.5 eV). Whilst our experiment corroborates a significant deviation from an idealized J$_{eff}$ = 1/2 magnetic ion, the large localized ordered moment found on the Ir site supports an antiferromagnetic Mott insulating state, rather than a metallic state that would emerge without large spin-orbit coupling in the presence of weak electron correlations.

\begin{center}
\begin{table*}
	\begin{tabular}{*{5}{c}}
		\hline \hline
		Magnetic & Ni$_1$ & Ni$_2$ & Ir$_1$ & Ir$_2$ \\
		sites & $(0, 0, \frac{1}{4})$ & $(0, 0, \frac{3}{4})$ & $(0, 0, 0)$ & $(0, 0, \frac{1}{2})$\\
		\hline 
 		$\tau_1$ & $(0, 0, u_1)$ & $(0, 0, -u_1)$ & $(0, 0, u_2)$ & $(0, 0, -u_2)$ \\[2ex]
		$\tau_2$ & $(0, 0, u_1)$ & $(0, 0, u_1)$ & $(0, 0, u_2)$ & $(0, 0, u_2)$ \\[2ex]
		$\tau_3$ & $ \begin{matrix} \frac{3}{2}(u_1+p_1, 0, 0) \ + \\ i\sqrt{3}(\frac{1}{2}(-u_1+p_1),-u_1+p_1, 0) \end{matrix} $ & $ \begin{matrix} \frac{3}{2}(0, v_1+w_1, 0) \ + \\ i\sqrt{3}(v_1-w_1,\frac{1}{2}(v_1-w_1), 0) \end{matrix} $ & $ \begin{matrix} \frac{3}{2}(u_2+p_2, 0, 0) \  + \\ i\sqrt{3}(\frac{1}{2}(-u_2+p_2),-u_2+p_2, 0) \end{matrix} $ & $ \begin{matrix} \frac{3}{2}(0, v_2+w_2, 0) \ + \\ i\sqrt{3}(v_2-w_2,\frac{1}{2}(v_2-w_2), 0) \end{matrix} $  \\
		\hline \hline
	\end{tabular}
	\caption{Magnetic configurations associated with the three irreducible representations of the group of the propagation vector {\bf k} = (0,~0,~1). $u_1$, $u_2$, $p_1$, $p_2$, $v_1$, $v_2$, $w_1$ and $w_2$ are refinable parameters.}
	\label{tab:mag-struct}
\end{table*}
\end{center}

\begin{center}
\begin{table*}
	\begin{tabular}{{c|cccccc}}
		\hline \hline
		 IR & \multicolumn{6}{c}{Symmetry generators of space group $R\bar{3}c1'$} \\
		  & 3$^+$(0,0,$z$) & 2($x$,$x$,0) & $\bar{1}$ & {\bf t$_1$} & {\bf t$_2$}& 1'  \\
		\hline 
		LD1 & $\begin{pmatrix} 1 & 0 \\ 0 & 1 \end{pmatrix}$ & $\begin{pmatrix} 0 & 1 \\ 1 & 0 \end{pmatrix}$ & $\begin{pmatrix} 0 & e^{i\pi k} \\ e^{-i\pi k} & 0 \end{pmatrix}$ & $\begin{pmatrix} e^{i\frac{2\pi k}{3}} & 0 \\ 0 & e^{-i\frac{2\pi k}{3}} \end{pmatrix}$ & $\begin{pmatrix} e^{i\frac{4\pi k}{3}} & 0 \\ 0 & e^{-i\frac{4\pi k}{3}} \end{pmatrix}$ & $\begin{pmatrix} -1 & 0 \\ 0 & -1 \end{pmatrix}$\\
		M$_z$ & M$_z$ & -M$_z$ & M$_z$ & M$_z$ & M$_z$ & -M$_z$ \\ 
		\hline \hline
	\end{tabular}
	\caption{Matrix representatives \cite{Aroyo2006} of the LD1 irreducible representation for generators  of the space group $R\bar{3}c1'$ (Miller and Love notation \cite{Cracknell1979}) and wave-vector {\bf k} =(0,0,$k$). {\bf t$_1$} and {\bf t$_2$} represents the respective R-lattice translations (2/3,~1/3/,~1/3) and (1/3,~2/3,~2/3).  The 1' symbol represents time-reversal symmetry. The bottom line shows how the M$_z$ secondary ferromagnetic mode at k=0 transforms under the various symmetries.}
	\label{tab:IR}
\end{table*}
\end{center}

\section{Discussion}

\par The magnetization measurements show two characteristic temperatures, $T_2 = 17\ \kelvin$ and $T_1 = 75\ \kelvin$, similarly to what have been reported in the studies made on polycrystalline samples \cite{Nguyen1995,Flahaut2003,Mikhailova2012}. The first one corresponds to a drop of the ZFC susceptibility, the second one to a maximum in the {\bf H} $\parallel$ {\bf c} magnetization, to the rise of magnetic reflections in neutron diffraction and to the onset of spin dynamics. AC susceptibility measurements show an out-of-phase signal whose maximum shifts towards higher temperatures when the frequency increases \cite{Costes2005}. We performed the same measurements on polycrystalline sample and found that the frequency shift follows an Arrhenius law : $\tau = \frac{1}{2\pi f} = \tau_0\cdot \exp(\frac{E}{k_B T})$, where $\tau$ corresponds to the relaxation time for which $\chi''(T)$ is maximum. We found an energy barrier E of $377\ \kelvin $ which is sightly lower than the value of $452.3~\kelvin$ obtained by Costes {\it et al.} \cite{Costes2005}. The peak in $\chi''(T)$ is observed between $T_1$ and $T_2$ and disappears below $T_2$ which indicates that the observed spin dynamic is associated to the transition occurring at $T_1$ and becomes too slow below $T_2$ to be observed at the time scale of the AC susceptibility measurement.

\par Two models were proposed to account for the magnetic behavior of \SNIO\ common to the other members of the family such as Sr$_3$CoIrO$_6$, Sr$_3$NiRhO$_6$ and Ca$_3$CoRhO$_6$. 
\begin{itemize}
\item A partially disordered antiferromagnetic state where 2/3 of the chains are antiferromagnetically ordered while 1/3 remains disordered between $T_1$ and $T_2$. Below $T_2$, the difference between the ZFC and FC susceptibility is explained by a freezing of the spin dynamics associated to the disordered chains producing a spin-glass like transition. The neutron diffraction pattern is well refined within this model with a ${\bf k} = (0, 0, 1)$ and taking in account a magnetic phase $\varphi = \frac{\pi}{6}$.
\item An amplitude modulated antiferromagnetic arrangement of ferrimagnetic chains as the one shown in Fig.~\ref{fig:mag-s}. In this case the ZFC-FC branching away is attributed to the dynamics of domain walls along the chains. This configuration accounts equally well for the neutron data at low temperature using a propagation vector ${\bf k} = (0, 0, 1)$ and a magnetic phase $\varphi = 0$.
\end{itemize}

\par Within the partially disordered antiferromagnetic model, the spin dynamics is related to the disordered chains and the step-like behavior observed at 15 $\kelvin$ can be easily explained. Without any applied magnetic field 1/3 of the chains are disordered while the remaining 2/3 are antiferromagnetically coupled. When a small magnetic field is applied, the magnetic moments of the disordered chains align along the direction of the magnetic field yielding a magnetization value of 1/3 of the saturated magnetization of the ferrimagnetic chains. 
\par In the case of the amplitude modulated antiferromagnetic model, all the chains are antiferromagnetically ordered and all the magnetic moments are compensated. The spin dynamics could be attributed to the nucleation of domain walls along the chains. The energy barrier determined from the AC susceptibility measurement is thus the energy needed in order to create a domain wall involving an anisotropy barrier and intrachain interactions \cite{Lhotel2008}.
Our neutron diffraction study combined with symmetry analysis explaining the origin of the weak ferromagnetic component clearly favours the amplitude modulated model at low temperature. We are unable to conclude however on the exact origin of the magnetic state above T$_2$ and its impact on the spin dynamics. We note that irrespective of the chain magnetic building blocks (antiferromagnetic or ferromagnetic intrachain coupling, spin and orbital contributions of the two chemical species), a universal behavior driven by magnetic frustration is evidenced in this class of materials, with a step in the magnetic isotherms and two anomalies in the magnetic susceptibility, respectively associated with magnetic ordering and a freezing of the spin dynamics, and . Finally, it is worth recalling the metastability of the magnetic structure in Ca$_3$Co$_2$O$_6$ \cite{Agrestini2011}. A similar metastability might also occur in the other members of the family which could be of interest to check in long-time scale neutron experiments. The appearance of a second magnetic phase might also be a possible explanation for the presence of the weak ferromagnetic component observed below 17~$\kelvin$ if this phase were ferro or ferrimagnetic. We cannot rule out this alternative explanation although we have no evidence for this at the moment from our neutron diffraction results.


\section{Conclusion}

\par In conclusion, we have clarified the magnetic behavior of \SNIO\ through single crystal magnetization study and neutron powder diffraction. We have evidenced an easy axis of anisotropy along the chain axis by single crystal magnetization measurements. We have established the occurrence of a magnetic order at the temperature $T_1$ at which the magnetic susceptibility ceases to show a Curie-Weiss behavior and determined the magnetic configuration up to an arbitrary global phase factor in the high temperature regime. At low temperature, a symmetry analysis of the antiferromagnetic order parameter shows that the onset of weak ferromagnetism along c is only compatible with a commensurate magnetic state with wave-vector {\bf k}=(0, 0, 1), and a specific overall phase of the modulation. Our study reveals sizeable ordered moments on the Ni and Ir sites, supporting a Mott insulating state. The divergence of magnetic susceptibility under field-cooled and zero-field cooled conditions at $T_2$ is associated to a spin freezing and a change of the magnetic state, either becoming incommensurate or associated to a modification of the overall phase. Our study lifts some inconsistencies reported in the literature and confirms the universality of the remarkable properties in this family of materials, despite the strong spin-orbit coupling of the iridium ion.


\section*{Acknowledgment}

\par We thank A.D. {\scshape Hillier}, P. {\scshape Manuel} and W. {\scshape Kockelmann} for their involvement in the project. We warmly thanks E. {\scshape Lhotel} for fruitful discussion and A. {\scshape Hadj-Azzem} for his collaboration on the synthesis. D.T.A. acknowledge financial assistance from CMPC-STFC grant number CMPC-09108.



\end{document}